# Absolute-Magnitude Calibration for W UMa-type Systems. II. Influence of Metallicity

Slavek Rucinski[1]

81 Longbow Drive, Scarborough, Ontario M1W 2W6, Canada

March 1, 1995

## ABSTRACT

A modification to the absolute magnitude calibration for W UMa-type systems, taking into account differences in metal abundances, is derived on the basis of contact binary systems recently discovered in metal-poor clusters. A preliminary estimate of the magnitude of the metallicity-dependent term for the $(B - V)$-based calibration is $\Delta M_V = -(0.3 \pm 0.1) \times [Fe/H]$. The calibration based on the $(V - I_C)$ color is expected to be less sensitive with the correction term $\approx -(0.12 \pm 0.05) \times [Fe/H]$.

## 1. Need for a metallicity-dependent term in the calibration

Searches for gravitational micro-lenses are currently giving large numbers of serendipitously discovered variable stars. Among these variables, there are many W UMa-type contact binaries. On the basis of the first part of the catalog of variable stars discovered during the OGLE project (Udalski et al. 1994), one can estimate the total number of contact binaries which will be discovered in the Baade Window during OGLE at well over one thousand. This estimate is based on 77 discoveries of such systems in the first part of the catalog which covered one of the 21 OGLE fields (i.e. less than 5% of the whole area searched), the Central Baade Window (BWC), and reached $I = 18$. The newly discovered systems will provide excellent statistics for the period, color and amplitude distributions of contact binaries, much better than those based on the sky-field sample which is heavily biased towards large-amplitude variables (Kaluzny & Rucinski 1993, Rucinski & Kaluzny 1994). They can be also utilized as calibrators of distances into the inner galactic disk (Rucinski 1995) with their absolute magnitudes estimated using the period-color calibration derived recently by Rucinski (1994a = CAL). The large numbers of contact binaries which are now being discovered should be compared with 562 W UMa-type systems in the most recent General Catalogue of Variable Stars. Among those 562, only 130 systems have good-quality data, but only for half of that in standard photometric systems.

Stars visible toward the Galactic Bulge show a large spread in metallicity. While the Bulge stars might show a range of metallicities as large as $-1 < [Fe/H] < +1$ (Frogel 1988), all contact

---

[1] Affiliated with the Department of Astronomy, University of Toronto and Department of Physics and Astronomy, York University

binary systems visible in BWC (Rucinski 1995) are at distances $< 6$ kpc from us and thus belong to the Disk population. For this population, the range in metallicities is expected to be $-0.5 < [Fe/H] < +0.5$.

The calibration presented in CAL utilized the observational (de-reddened) $B - V$ color as the temperature index. Since red-based color indices should become more popular in the future, and the OGLE project provides data in $V$ and $I_C$ spectral bands, an additional, very preliminary calibration based on the $V - I_C$ color was also given in CAL. This calibration was derived assuming that the relation between the $B - V$ and $V - I_C$ colors is identical to that for the Main Sequence stars, an assumption which does not necessarily have to be correct for the W UMa-type systems because of their strong chromospheric activity. The relations given in CAL for predicting the absolute magnitudes $M_V$ at maximum light, together with an independent calibration for $M_I$ (Rucinski 1995) are:

$$M_V = -2.38 \log P + 4.26 \, (B - V) + 0.28, \quad \sigma = 0.24 \qquad (1)$$

$$M_V = -4.43 \log P + 3.63 \, (V - I_C) - 0.31, \quad \sigma = 0.29 \qquad (2)$$

$$M_I = -3.23 \log P + 2.88 \, (V - I_C) - 0.03, \quad \sigma = 0.29 \qquad (3)$$

The period $P$ is in days and both colors are reddening-free. The predictive power of these relations was estimated in CAL to be at a level of about $\pm 0.5$ mag. Extensive Monte-Carlo experiments show that the uncertainty depends on a particular period–color combination and is smaller (typically $\pm 0.2$ mag) within the strict range of periods and colors used to derive the calibrations ($0.27 < P < 0.59$ day, $0.40 < B - V < 1.08$, $0.46 < V - I_C < 1.19$).

The metallicity dependence has been by-passed in CAL as the calibration was based on a few nearby W UMa systems which presumably have solar abundances and on four nearby open clusters with moderate under-abundances ($-0.16 < [Fe/H] < 0$, see Table 1 in CAL). The $(B - V)$-based version of the calibration was applied also in CAL to three distant clusters with a larger range of metallicities. In this case, the goal was to sieve out non-member, foreground and background, Milky Way interlopers from these low galactic-latitude clusters. Mazur et al. (1995) applied recently the calibration for the same purpose to establish membership of 18 among 28 W UMa-type systems in the field of Cr 261, and Yan & Mateo (1994) used it for 4 systems in the globular cluster M71 (see below).

The three low-latitude clusters analyzed in CAL had $[Fe/H]$ between $-0.6$ and $+0.19$. It was noted that the most metal-poor cluster To 2 showed consistently positive deviations $\Delta M_V$ from the calibration, $\Delta M_V = M_V^{obs} - M_V^{pred}$ (where $M_V^{obs}$ is derived simply from the brightness of a system at maximum light and the assumed distance modulus of the cluster). The metallicity of To 2 is definitely low but is poorly known at this moment; it might be as low as $[Fe/H] = -1.2$ (Geisler 1987). We would like to establish if the deviations observed for To 2 could be caused by the low metallicity of the cluster.

This paper addresses the metallicity dependence of the absolute magnitude calibrations given by Equations (1) and (2), in view of recent discoveries of W UMa-type binaries in metal-poor globular clusters.

## 2. Determination of the correction on the basis of metal-poor globular clusters

Little is known about metal abundances of isolated W UMa systems, mostly because of the extreme rotational blending of spectral lines which prevents application of traditional techniques. Modern methods of spectral synthesis could possibly shed light on this matter but have not been yet applied. *ubvy* photometry of a sample of bright W UMa systems (Rucinski & Kaluzny 1981, Rucinski 1983) has not lead to clear-cut results as it was impossible to exclude the presence of intrinsic, activity-related peculiarities in ultraviolet parts of spectra. Fortunately, recent discoveries of contact systems in very metal-poor stellar systems offer a possibility of a preliminary look into the dependence of absolute magnitudes on metal abundances.

Modifications to the *period–color* relation resulting from metallicity differences were already discussed by Rucinski (1994b = POP2). This study concentrated on limits imposed by the full-convection condition for Population II stars but implicitly discussed observed modifications to periods and colors for low metallicity systems. It was shown there that two systems in NGC 5466, NH 19 and NH 30 (Mateo et al. 1990), and four systems in NGC 4372 (Kaluzny & Krzeminski 1993), V4, V5, V16 and V22, are bluer and have shorter periods than the common, Population I W UMa systems. Application of the $M_V = M_V(\log P, B-V)$ calibration confirmed that these systems were slightly fainter than expected, as should be for Population II stars. Thus, six *bona fide* representatives of metal-poor populations were identified at that time. It was argued in POP2 that the remaining 4 systems in NGC 4372 were Milky Way foreground interlopers.

It should be stressed that data for the 6 contact systems in both of the metal-poorest clusters, NGC 5466 and NGC 4372, are of mediocre quality due to difficulties of observing faint variable stars. However, both clusters are so metal-poor that even crude data could give information on major trends due to metallicity variations.

Recently, Yan & Mateo (1994) discovered 4 contact binaries in M71 increasing the sample of W UMa-type systems in the globular clusters from 6 to 10. This is a very important discovery as the relatively moderate metal under-abundance of the cluster, estimated at about $[Fe/H] = -0.7$, permits to bridge the large gap between the solar-abundance systems and the metal-poor ones. The data were obtained in the $V$ and $I_C$ filters so that Eq. (2) must be used to calculate $M_V^{pred}$ for systems in M71. This forces us to discuss together results based on the $B-V$ color with results based on the $V-I_C$ color. We address the specific question of relative sensitivities of both colors to metallicities in the last section of the paper.

The assumptions made to determine the absolute magnitudes $M_V$ and the reddening-corrected colors of systems in the three globular clusters were as follows. For NGC 5466, $m - M = 16.0$, $E(B-V) = 0$; for NGC 4372, $m - M = 14.8$, $E(B-V) = 0.48$ (with additional corrections for patchy extinction, as described by Kaluzny & Krzeminski 1993); and for M71, $m - M = 13.7$ and $E(B-V) = 0.28$. The values of periods and colors for the W UMa systems were taken from the discovery papers and are listed in Table 1.

Figure 1 shows the deviations $\Delta M_V$ from the Population I calibration as function of $[Fe/H]$ for 18 systems in stellar systems of low metal abundance. These are the 14 systems in three globular clusters (4 new systems M71 and 10 systems as in POP2) and the 4 systems in the open cluster To 2 (as in CAL). The values of $[Fe/H]$ for these clusters were taken from Geisler (1987) and Geisler et al. (1992): $-1.2$ for To 2, $-2.22$ for NGC 5466, and $-2.35$ for NGC 4372; $[Fe/H] = -0.7$ was assumed for M71. Less extreme values of $[Fe/H]$ have been also given in the literature: for To 2, Kubiak et al. (1992) assumed $[Fe/H] = -0.6$ and the database of Webbink (1985) gave $-1.85$ and $-1.77$ for NGC 5466 and NGC 4372, respectively. The systems which we consider as members of the clusters have been marked in Figure 1 by filled symbols. The broken lines give the approximate range for the membership acceptance/rejection for Population I systems, estimated at $\pm 0.5$ mag in CAL. On the basis of Figure 1, we conclude:

1. The globular cluster data suggest a weak dependence of the absolute magnitude $M_V$ on $[Fe/H]$; a linear relation should suffice for the relatively poor data which are available now.

2. Inclusion or rejection of M71 would have only a minor influence on the slope of a linear relation, as this cluster provides a too small baseline in $[Fe/H]$. Besides, the deviations for this cluster were derived from the calibration based on $V - I_C$, and not on $B - V$, as for the remaining clusters. As we show in Section 3, the metallicity dependence entering through the color term is expected to be different for these two color indices.

3. Assuming that the dependence is linear and using the data for the 6 metal-poor globular cluster systems with metallicities based on Geisler et al. publications, we obtain the slope $\Delta M_V / [Fe/H] = -0.26 \pm 0.12$; for the set of less-extreme metallicity data, as mentioned above, the slope is $-0.34 \pm 0.12$.

4. The four systems in To 2 deviate too much from the linear relation shown in Figure 1 to maintain that their under-luminosity is due to low metallicity; the most likely reason for the deviations is the poorly known reddening of the cluster (Kubiak et al. 1992).

5. It is suggested that until additional data are available, the calibration based on the observed $B - V$ color (Eq.(1)) should be augmented by a term $-(0.3 \pm 0.1) \times [Fe/H]$. We do not have sufficient data to determine observationally the coefficient for the calibration based on $V - I_C$ but, on the basis of the model atmospheres (see next section), we estimate that the dependence should be about 2.5 times weaker.

We note that the new discovery for M71 by Yan & Mateo for M71 has resulted in an identification of *the first known contact binary systems which are located below the turn-off point* *of a globular cluster*. This should be contrasted with our calibrating systems, *all six of them being Blue Stragglers*, as is clearly visible in Figure 2 which shows the 14 globular-cluster systems in the color-magnitude diagram. We are left, as in POP2, with only the Blue Stragglers systems when we reject binaries which (1) probably are foreground interlopers as they are much too bright ($> 1$ mag) relative to the $M_V$ calibration (see Figure 1) and (2) have period-color combinations such as for Population I systems. At the time of writing POP2 we were confident that this way one can effectively remove Population I field interlopers from the sample (such systems are marked by open symbols in Figures 1 and 2). In the case of the systems in M71, this approach encounters a complication: While the agreement of absolute magnitudes is basically perfect, the period-color relation does not indicate any obvious blue deviations. As we can see in Figure 3, all four systems fall into the main band defined by sky-field, presumably old-disk (Guinan & Bradstreet 1988) W UMa-type systems. We do not have a ready explanation for this discrepancy. Possibly, $[Fe/H] = -0.7$ is too-mild a deficiency to affect the colors, or the transformation of $V - I_C$ to $B - V$ is not applicable here, or – and this would be most unfortunate – the 6 systems used in this paper to find the $[Fe/H]$ dependence are not representative for the whole population because of their Blue Straggler status.

## 3. The metallicity dependence of the color term

The previous section gives us some idea about a correction to the absolute magnitude calibrations based on the $B - V$ color, but we have no observational data to establish how sensitive to metallicity is the calibration based on the $V - I_C$ color. We should stress that it is a low-order simplification to describe properties of contact binaries by only two terms in the calibration, $\log P$ and the color (cf. derivations of the expected relationships in CAL). But, if we really want to use very simple terms, we can expect that when metallicity of a contact system is varied, its period will be modified according to changes in radii driven by internal structure modifications while its color will change through changes in the atmospheric properties. The situation is reminiscent to that for subdwarfs which are known to be less luminous than Population I stars partly because of smaller radii, and partly because large effects of metallicity on their spectra and colors.

To place the result of the previous section in a predictive context, we can *assume* that the metallicity modification to our calibration is entirely due to the color term. This dependence is much easier to predict than the modification to the orbital periods which would require calculation of detailed structure models, with several unavoidable assumptions on the way. Whereas the metallicity dependence of $U - B$ and $B - V$ colors for solar-type stars has attained a textbook-case status, similar inter-comparisons involving the $V - I_C$ color are rare. A convenient tabulation has been recently provided by Buser & Kurucz (1992) who computed model atmosphere colors for several combinations of parameters for solar-type stars with $4000 \leq T_{eff} \leq 6000$ K. We used only the metallicity dependence of these theoretical models and disregarded the still-existing problems of matching the theoretical colors to the observed ones. The differences between these colors for various values of metallicities $[Fe/H]$ and for the solar metallicity are shown in Figures 4 and 5. On the basis of these figures, we conclude the following:

1. For the range of moderate deviations from the solar metallicity, $-0.5 < [Fe/H] < +0.5$, the corrections to colors scale approximately as $\Delta(B - V) \propto +0.08 \times [Fe/H]$ and $\Delta(V - I_C) \propto +0.04 \times [Fe/H]$.

2. There exists practically no dependence on the gravity, but there is some dependence on the stellar effective temperature, with the $\Delta(V - I_C)$ color-deviation relatively more dependent on it.

3. For strong metal under-abundances, the relations tend to "saturate" and become less sensitive to $[Fe/H]$. For $[Fe/H] < -2$, the corrections become $\Delta(B - V) \simeq -0.15$ and $\Delta(V - I_C) \simeq -0.05$.

With these model results, we are in position to check consistency of the assumption that the metallicity correction to the $M_V(\log P, B - V)$ calibration is dominated by the color term. Comparing the color-term coefficient 4.26 in Eq. (1), with the approximate slope of the $B - V$ metallicity dependence described above, $0.08 \times [Fe/H]$, we indeed approximately recover the correction term $\propto 0.3 \times [Fe/H]$. This close agreement must be partly fortuitous, because the data are not accurate, but it does confirm that the the metallicity enters mainly through the color term of the calibration.

We can now apply the same argument to the calibration based on $V - I_C$ (Eq.(2)). The color term of this calibration is smaller and the $V - I_C$ color is about 2 times less sensitive to changes in metallicity than the $B - V$ color. Thus, we guess at this point that the correction to the $M_V(\log P, V - I_C)$ should include a metallicity term whose magnitude is $\approx 2.5\times$ smaller than for the $(B - V)$-based calibration, i.e. $-(0.12 \pm 0.05) \times [Fe/H]$. This is a very preliminary estimate requiring an observational confirmation.

## 4. Conclusions

The correction term to the $M_V$ calibration based on the observational, de-reddened color $B - V$, $\Delta M_V \simeq -0.3\,[Fe/H]$, has been obtained on the basis of 6 W UMa-type systems which are certain members of two very metal-poor globular clusters. All 6 systems are Blue Stragglers which inserts a degree of uncertainty into applicability of the present results to all observable contact-binary systems. Comparison of the metallicity correction $\Delta M_V$ with the model-atmosphere results on synthetic colors suggests that the largest effect in the calibration is through the color term. This permits us to formulate a prediction that the absolute magnitude calibration based on the $V - I_C$ color is about 2.5 times less sensitive to metallicity variations than the one based on the $B - V$ color.

Thanks are due to my wife Anna for her support of my research. The research grant from the Natural Sciences and Engineering Council of Canada is acknowledged with gratitude.

---

TABLE 1

W UMa-type Systems in Globular Clusters

| System | P (d) | $(B-V)_0$ | $M_V^{obs}$ | $\Delta M_V$ | Comment |
|---|---|---|---|---|---|
| **NGC 5466** | | | | | $[Fe/H] = -2.22$ |
| NH 19 | 0.3421 | 0.15 | 2.42 | +0.39 | |
| NH 30 | 0.2975 | 0.155 | 3.23 | +1.06 | |
| | | | | | |
| **NGC 4372** | | | | | $[Fe/H] = -2.35$ |
| V4 | 0.2418 | 0.35 | 4.08 | +0.84 | |
| V5 | 0.3403 | (0.35) | 4.27 | +1.39 | $(B-V)_0$ from $(V-I_C)_0$ |
| V16 | 0.3084 | 0.33 | 3.04 | +0.14 | |
| V22 | 0.4150 | 0.15 | 2.04 | +0.21 | |
| V27 | 0.3772 | 0.98 | 2.68 | −2.78 | non-member |
| V31 | 0.3777 | 0.64 | 2.67 | −1.34 | non-member |
| V33 | 0.2791 | 0.90 | 4.29 | −1.14 | non-member |
| V35 | 0.2901 | 0.71 | 2.60 | −1.98 | non-member |
| | | | | | |
| **M71** | | $(V-I)_0$ | | | $[Fe/H] = -0.7$ |
| V1 | 0.3489 | 0.75 | 4.59 | +0.15 | |
| V2 | 0.3672 | 0.60 | 4.11 | +0.31 | |
| V3 | 0.3739 | 1.06 | 5.37 | −0.06 | |
| V5 | 0.4045 | 0.64 | 4.24 | +0.51 | |

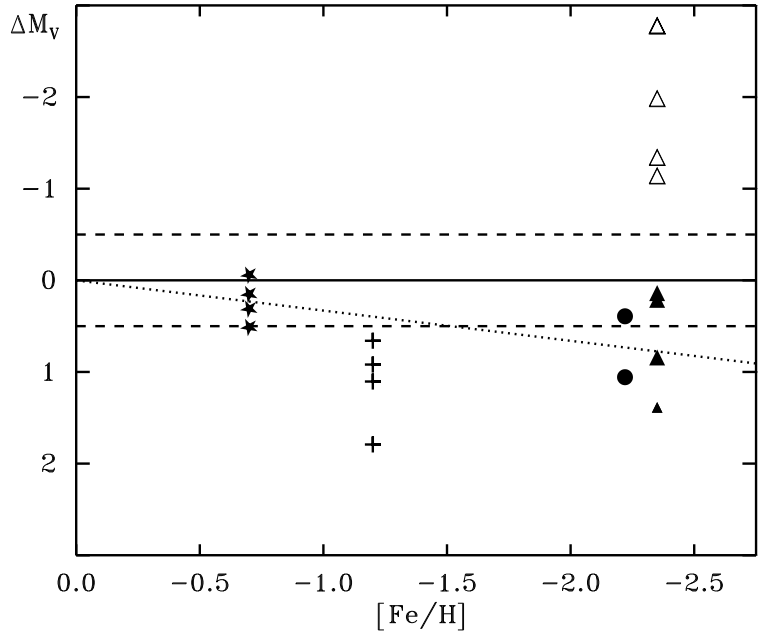

Fig. 1.— Deviations from the $M_V$ calibration for 14 W UMa-type systems in 4 metal-poor clusters: the open cluster To 2 (crosses) and the globular clusters NGC 5466 (circles), NGC 4372 (triangles) and M71 (asterisks). The systems which we think are members of globular clusters are marked by filled symbols. Note that two different $M_V$ calibrations were used: for most systems, it was the one based on $B - V$ but, for M71, the one based on $V - I_C$ was used; it is assuring that these calibrations seem to give consistent results. V5 in NGC 4372 has been marked by a smaller symbol since its $B - V$ was not measured directly but was derived from $V - I_C$. The $[Fe/H]$ values for NGC 5466 and NGC 4372 are from the papers by Geisler et al. The dotted line gives the linear fit for 6 member systems in metal-poor globular clusters with the slope of $-0.3$.

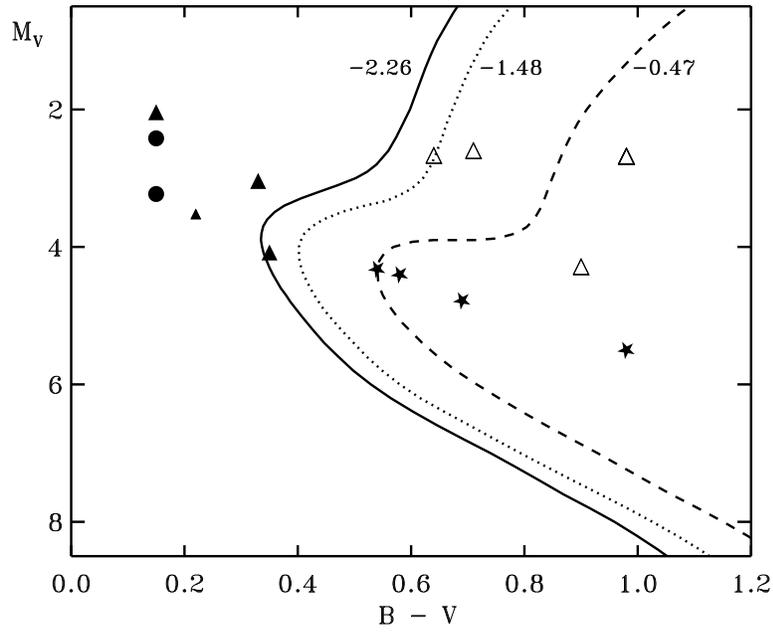

Fig. 2.— The same systems as in Figure 1 are shown in a combined color – magnitude diagram. The isochrones for $14 \times 10^9$ years and for 3 values of $[Fe/H]$ (as labeled) are from Bergbusch & VandenBerg (1992). For M71, the $B - V$ values were derived from the de-reddened $V - I_C$ colors using the Main Sequence relation.

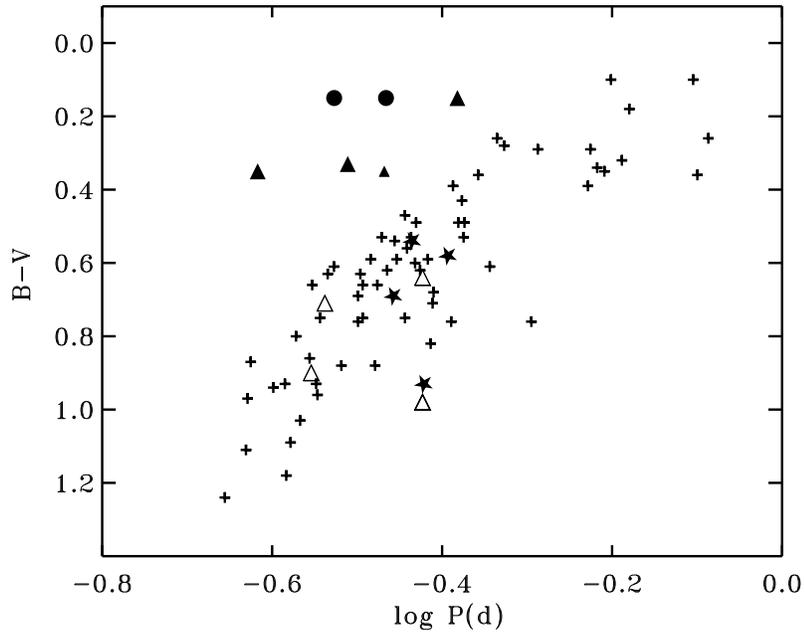

Fig. 3.— The period-color for bright W UMa-type systems from the compilation of Mochnacki (1985) (crosses) and for systems in globular clusters (same symbols as in Figures 1 and 2, but the data points for To 2 are not plotted). Note that the systems in M71 (asterisks) do not have abnormally blue colors.

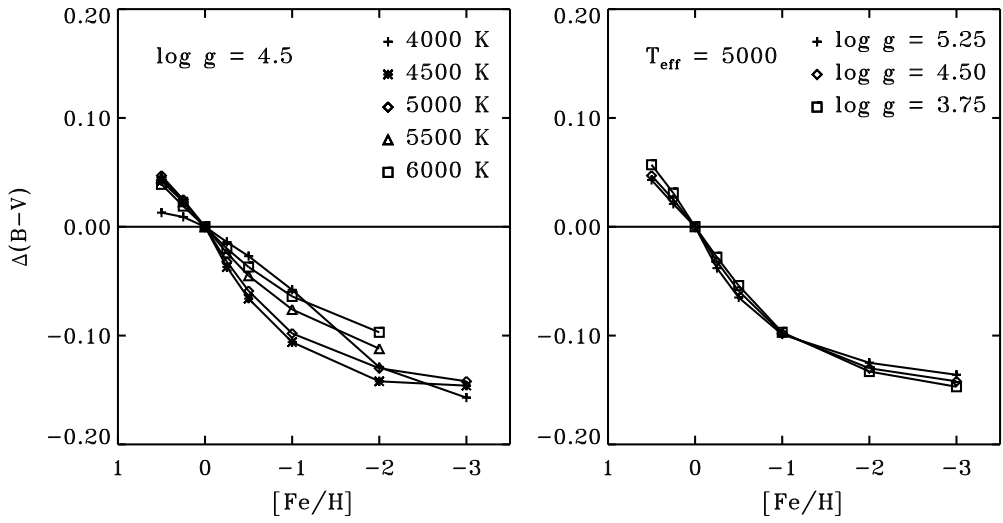

Fig. 4.— Dependence of $B - V$ on $[Fe/H]$ for solar-type stars, following the model atmosphere results of Buser & Kurucz (1992).

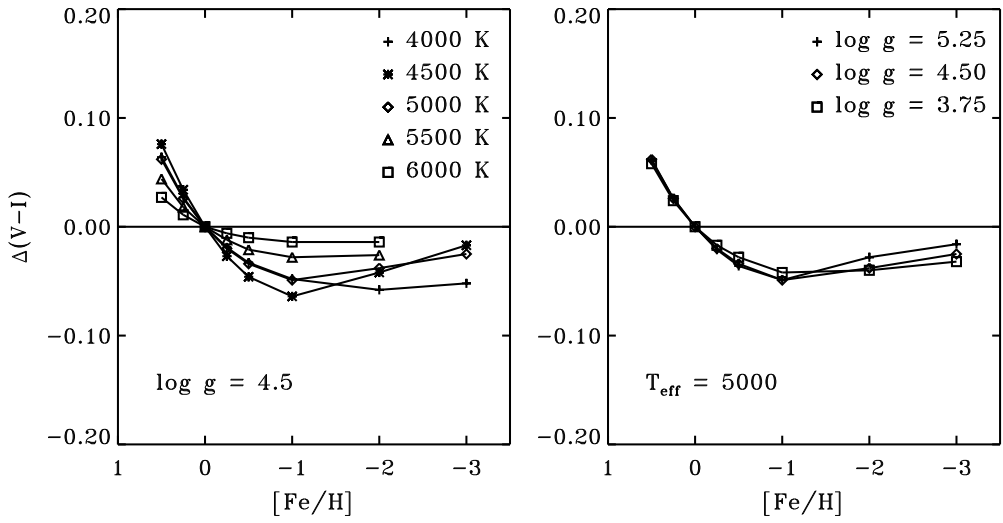

Fig. 5.— Dependence of $V - I_C$ on $[Fe/H]$ for solar-type stars, following the model atmosphere results of Buser & Kurucz (1992).